\documentclass[12pt]{article}
\def\be{\begin{equation}}
\def\ee{\end{equation}}
\def\ba{\begin{array}{c}}
\def\ea{\end{array}}

\def\ben{\[ }
\def\een{\] }

    \def\be{\begin{equation}}

\begin{document}
\title{Solvable relativistic quantum dots with vibrational
spectra
%\,\footnote{For footnotes to
 %      the title, authors, and addresses, the footnotemarks
  %     [$^*),\ ^{**}),\ ^{\dagger})$, etc.] are used
   %    automatically.}
       }
\author{Miloslav Znojil}

\date{\'{U}stav jadern\'e fyziky AV \v{C}R, 250 68 \v{R}e\v{z},
Czech Republic}

%\pacs{03.65.Ge} \keywords{Klein-Gordon equation, point
%interactions, PT symmetry}
%%%%%%%%%%%%%% FOR EDITORIAL USE ONLY!!! %%%%%%%%%%%%%%%

%\lastpage{000}
%\makefirsttitle
%%%%%%%%%%%%%%%%%%%%%%%%%%%%%%%%%%%%%%%%%%%%%%%%%%%%%%%%
\maketitle
%Page headings:
\begin{abstract}

For Klein-Gordon equation a consistent physical interpretation of
wave functions is reviewed as based on a proper modification of
the scalar product in Hilbert space. Bound states are then studied
in a deep-square-well model where spectrum is roughly equidistant
and where a fine-tuning of the levels is mediated by ${\cal
PT}-$symmetric interactions (composed of imaginary delta
functions) which mimic creation/annihilation processes.

\end{abstract}

\newpage

\section{Klein-Gordon equation \label{prela}}

\subsection{Pseudo-Hermitian Feshbach-Villars Hamiltonian
 \label{prelua}}

As long as the most common relativistic  Klein-Gordon (KG)
operators are partial differential operators of the second order
with respect to time, the time evolution of the wave functions
$\Psi^{(KG)}(x,t)$ must be studied together with their first time
derivatives $i\,\partial_t\,\Psi^{(KG)}(x,t)$. After the routine
Fourier transformation we arrive at the Feshbach-Villars (FV,
\cite{FV}) non-Hermitian eigenvalue problem
 \be
  \hat{H}^{(FV)} \,|\psi\rangle =
 E\,|\psi\rangle\,,
 \ \ \ \ \ \ \ \ \ \
 \hat{H}^{(FV)}=\left (
 \begin{array}{cc}
 0&\hat{h}^{(KG)}\\
 1&0
 \ea
 \right )
 \label{KGE}
 \label{jepiodnadfr}
 \ee
where the two wave-function components may be marked as $| D \}$
(=``down component") and $| U \}$ (=``up component"). For the
description of the bound states in one dimension the two-by-two
partitioning in (\ref{KGE}) allows us to extract $ |U\} =E\, |D\}$
and to replace our Klein-Gordon equation by its reduced form
 \be
 \hat{h}^{(KG)}\,|
 D_n\}=\varepsilon_n\, |D_n\},
 \ \ \ \ \ \
 n =  1,2, \ldots
 \,
 \label{SE}
 \label{upeddfr}
 \ee
with squared energy $E^2$ abbreviated as $\varepsilon$ and with
the ``large" Hilbert space ${\cal H}$ of kets $ |\psi\rangle$
reduced to the ``smaller" Hilbert space ${\cal H}_{(c)}$ of the
{\bf c}urly-ket ``down" components $ |D_n\}$~\cite{polub}.

\subsection{Biorthogonal bases}

The ``right" eigenkets $|D_n\}$ will not carry all information
about $ \hat{h}^{(KG)}$ whenever $[\hat{h}^{(KG)}]\neq
[\hat{h}^{(KG)}]^\dagger$. Then, the parallel Schr\"{o}dinger-type
problem generates {\em different} eigenkets marked by the double
curly ket symbol. The latter sequence may be re-read as the left
eigenvectors of our original operator $ \hat{h}^{(KG)}$, related
to the {\em same} (by assumption, real) eigenvalues $\varepsilon_n
\equiv \kappa_n^2$,
 \be
  \{ \{ L_n|\,\hat{h}^{(KG)}
 =\kappa^2_n\, \{ \{ L_n|
 , \ \ \ \ \ \ n =  1, 2,\ldots \,.
 \label{smaller}
 \ee
It is well known that the set of the bras $ \{ \{ L_n|$ and kets
$|D_n\}$ is bi-orthogonal \cite{polub},
 \ben
  \{ \{ L_m| D_n\} =
  \left \{
  \begin{array}{l}
  0 \\
  \varrho_{n}\neq 0  \ea
  \right .\ \ \ {\rm for} \ \ \ \ \left \{
  \begin{array}{l}
 m \neq n,\\
 m=n\,,
 \ea
  \right .\
  \een
and that it forms, usually, a basis in the infinite-dimensional
Hilbert space ${\cal H}_{(c)}$. Then, we may decompose the unit
operator and/or derive the bi-orthogonal spectral representation
of the Hamiltonian  in ${\cal H}_{(c)}$,
 \be
 I_{(c)} = \sum_{n=1}^\infty\,|D_n\}\,
 \frac{1}{\varrho_n}\,\{ \{ L_n|\,,\ \ \ \ \ \ \ \
 \hat{h}^{(KG)}=
 \sum_{n=1}^\infty\,|D_n\}\,
 \frac{\kappa_n^2}{\varrho_n}\,\{ \{ L_n|\,.
 \label{selike}
 \ee
The overlaps $\varrho_n$ need not be all of the same sign.

\section{Relativistic observables}

\subsection{$\Theta-$quasi-Hermiticity
 \label{prelubec}}

In the space  ${\cal H}={\cal H}_{(c)}\oplus {\cal H}_{(c)}$ of
the  eigenstates of $H^{(FV)}$ we have to consider the pair of
conjugate equations
 \be
 \hat{H}^{(FV)}\,|n^{(\pm)} \rangle =
 \pm \kappa_n\,|n^{(\pm)} \rangle , \ \ \ \ \ \ \ \
 \langle \langle n^{(\pm)} |\,
 \hat{H}^{(FV)} =
 \pm \kappa_n\,
 \langle \langle
 n^{(\pm)} | \,.
 \label{bigger}
 \ee
Both the left and right eigenstates have the two-component
structure,
 \ben
 |m^{(\pm)} \rangle \rangle =
 \left (
 \ba
  |L_m\}\}\\ \, \pm {\kappa_m}\,|L_m\}\}
 \ea
 \right )
 \,, \ \ \ \ \ \ \ \
 |n^{(\pm)} \rangle =
 \left (
 \ba
 \pm {\kappa_n}\, |D_n\}\\ \, |D_n\}
 \ea
 \right )\,
 %\label{generated}
 \een
and form the bi-orthogonal set in the ``bigger" space  ${\cal H}$,
 \ben
 \langle \langle m^{(\nu)}
  | n^{(\nu\prime)}\rangle =
  \delta_{mn}\delta_{\nu\nu\prime}\cdot \mu_m^{(\nu)},
   \ \ \ \ \ \
   \mu_m^{(\pm)}
  =
  \pm 2\kappa_m\,\varrho_m\,,
  \ \ \ \ \ \ \nu,\,\nu\prime = \pm 1\,.
 \een
It is expected to be complete and useful,
 \be
 I = \sum_{\tau = \pm 1}\sum_{n=1}^\infty\,| n^{(\tau)}\rangle
 \,\frac{1}{\mu_n^{(\tau)}}\, \langle \langle n^{(\tau)} | \,,
  \ee
 \ben
 H^{(FV)} = \sum_{\tau = \pm 1}\sum_{n=1}^\infty\,| n^{(\tau)}\rangle
 \,\frac{\tau\,\kappa_n}{\mu_n^{(\tau)}}\, \langle \langle n^{(\tau)} |
 =
 \sum_{n=1}^\infty\, \frac{
 \left (| n^{(+)}\rangle
 \, \langle \langle n^{(+)}
 |\right )+\left (
 | n^{(-)}\rangle
 \, \langle \langle n^{(-)}| \right )
 }{2\,\varrho_n}
  \,.
  \een
Let us now assume that at a given $\hat{H}^{(FV)}$, equation
 \be
 \left [\hat{H}^{(FV)} \right ]^\dagger
  = {\eta}\,\hat{H}^{(FV)}\,{\eta}^{-1}
 \label{neherimod}
 \ee
possesses a positive and Hermitian solution $\eta_+ = \Theta > 0$.
Such an operator may play the role of a {\em metric} and induces
the following specific scalar product in ${\cal H}$,
 \be
  \left (
 |\psi_1\rangle \odot |\psi_2\rangle
 \right )
 =
 \langle \psi_1|\, \Theta\,|\psi_2 \rangle
 =
 \langle \psi_1 |\psi_2\rangle_{(physical)}, \ \ \ \ \ \ \
 |\psi_1\rangle
 \in {\cal H},  \ \ \ \
 |\psi_2\rangle
 \in {\cal H}\,.
 \ee
This product generates the norm, $||\psi||=\sqrt{\langle \psi
|\psi\rangle_{(physical)}}$. In terms of the later product and
metric we may call all the operators $A$ with the property
$A^\dagger=\Theta\,A\,\Theta^{-1}$ {\em quasi-Hermitian} and treat
them as observables (see \cite{Geyer} for a deeper outline of some
more sophisticated mathematical details). Indeed, we have
 \be
\left (
 |\psi_1\rangle \odot |A\,\psi_2\rangle
 \right )
 \equiv
  \left (
 |A\,\psi_1\rangle \odot |\psi_2\rangle
 \right )\,
 \ee
so that the probabilistic expectation values $\langle \psi|\,A\,
|\psi\rangle_{(physical)}$ are mathematically unambigously
defined.

%\section{Outlook}

\subsection{Explicit constructions of the metric $\Theta$ }

Let us assume non-Hermiticity of the type $ \hat{h}^{(KG)} \neq
[\hat{h}^{(KG)}]^\dagger= {\cal P}\,\hat{h}^{(KG)}\,{\cal P}$ in
the smaller space ${\cal H}_{(c)}$ (here, ${\cal P}$ is operator
of parity). Then, a consistent {\em physical} meaning may still be
assigned to all the relativistic bound states, provided only that
in the bigger space ${\cal H}$ we find a suitable physical metric
$\Theta$. For this purpose we may employ the ansatz
 \ben
 \Theta=
 \sum_{\tau,\tau\prime = \pm 1}
 \sum_{m,n=1}^\infty\,|n^{(\tau)}\rangle\rangle\,M_{nm}^{(\tau \tau\prime)}
 \,\langle\langle m^{(\tau\prime)}|
 \,,
 \een
the backward insertion of which in (\ref{neherimod}) gives the
condition
 \ben \tau\,\kappa_n\,M_{nm}^{(\tau \tau\prime)} =
M_{nm}^{(\tau \tau\prime)} \, \tau\prime\,\kappa_m
 \een
with the set of solutions $ M_{nm}^{(\tau \tau\prime)} =
\omega^{(\tau)}_n \delta_{nm}\,\delta_{(\tau \tau\prime)}$
numbered by the free parameters $\vec{\omega}^{(\pm)}$. The
Hermiticity and positivity constraints restrict the freedom of the
choice of both the optional sequences  $\vec{\omega}^{(\pm)}$ to
the real and positive values, $\omega^{(\pm)}_n > 0$. {\it Vice
versa}, {\em any} choice of the latter two sequences defines an
eligible operator of the metric
 \be
 \Theta=\Theta_{\vec{\omega}^{(\pm)}}=
 \sum_{\tau = \pm 1}
 \sum_{n=1}^\infty\,|n^{(\tau)}\rangle\rangle\, \omega^{(\tau)}_n
 \,\langle\langle n^{(\tau)}|
 \,.
 \ee
Its inverse
 \be
 \Theta^{-1}=
 \sum_{\tau = \pm 1}
 \sum_{n=1}^\infty\,|n^{(\tau)}\rangle\,
 \frac{1}{\omega^{(\tau)}_n
 |\,\mu_{n}^{(\tau)}|\,^2 }
 \,\langle n^{(\tau)}|
 \,
 \ee
is similar. In terms of the metric $\Theta$, the formal
bound-state wave functions re-acquire the standard probabilistic
interpretation.

\section{Models with complex point interactions \label{2}}

In a way inspired by the success of several non-relativistic
studies of ${\cal PT}-$symmetric models with point interactions
\cite{Kurasov} and by the encouraging experience we made in our
paper~\cite{DW} we shall combine the infinitely deep square-well
real part of the potential [$V(x) = \infty$ for all $x \notin
(-1,1)$] with the following purely imaginary delta-function
formula for its remaining part,
 \be
 V(x) =\sum_{\ell=1}^{\cal L}
 \left [
 i\,\xi_{\ell}\,\delta \left (x-a_{\ell}\right )-
 i\,\xi_{\ell}\,\delta \left (x+a_{\ell}\right )
 \right ]
 \,, \ \ \ \ \ \ \ \ x \in (-1,1)\,,
 \label{potik}
 \ee
at real couplings $\xi_\ell$ and ordered points $0 < a_1 <a_2 <
\ldots < a_{{\cal L}-1}<a_{\cal L}<1$.

\subsection{Wave functions}

The key advantage of our $V(x)$ in (\ref{potik}) is that the
${\cal PT}-$symmetrically normalized coordinate representants
$\psi(x)=\psi^*(-x)$ of $|D\}$ in eq. (\ref{upeddfr}) remains
piecewise trigonometric. At each real and positive bound-state
energy $\varepsilon=\kappa^2$ we shall have
 \be
 \psi(x)= \left \{
 \begin{array}{ll}
  \psi_L^{({\cal L})}(x) =
 (\alpha_{\cal L} - i\,\beta_{\cal L})\,\sin \kappa(1+x)
 , \ \ & x \in (-1,-a_{\cal L}), \\
 \multicolumn{2}{l}{
 \psi_L^{(\ell)}(x) =
 (\alpha_\ell - i\,\beta_\ell)\,\sin \kappa(a_{\ell+1}+x)+
 (\gamma_\ell - i\,\delta_\ell)\,\cos \kappa(a_{\ell+1}+x),
  }\\
 & x \in
 (-a_{\ell+1},-a_{\ell}),\\
  \psi_C^{(0)}(x) =
 \mu\,\cos \kappa x + i\,\nu \,\sin \kappa x, \ \
  & x \in (-a_1,a_1), \\
 \multicolumn{2}{l}{
 \psi_R^{(\ell)}(x) =
 (\alpha_\ell + i\,\beta_\ell)\,\sin \kappa(a_{\ell+1}-x)+
 (\gamma_\ell + i\,\delta_\ell)\,\cos \kappa(a_{\ell+1}-x),
  }\\
   & x \in
 (a_{\ell},a_{\ell+1}), \\
 \psi_R^{({\cal L})}(x) =
 (\alpha_{\cal L} + i\,\beta_{\cal L})\,\sin \kappa(1-x),
  \ \ & x \in (a_{\cal L},1), \ \ \
   1 \leq \ell < {\cal L}.
 \ea
 \right .
 \label{ansatz}
 \ee
Its  differentiation as well as continuity conditions
 \be
 \begin{array}{c}
  \psi_L^{(\ell-1)}(-a_{\ell})
  =\psi_L^{(\ell)}(-a_{\ell}), \ \ \
 \ell ={\cal L},{\cal L}-1,\ldots, 2,\\
  \psi_C^{(0)}(-a_{1})= \psi_L^{(1)}(-a_{1}),\ \ \ \
   \ \ \ \ \ \ \ \ \ \ \
   \psi_R^{(1)}(a_1)=
  \psi_C^{(0)}(a_{1}),\\
  \psi_R^{(\ell+1)}(a_{\ell+1})=
  \psi_R^{(\ell)}(a_{\ell+1}) , \ \ \
 \ell =1, 2, \ldots,{\cal L}-1,\\
 \ea
 \label{spojit}
 \label{sesit}
 \ee
enter the definition of the action of the delta functions,
 \be
 \begin{array}{c}
 \left [
  \psi_L^{(\ell-1)}(-a_{\ell})
  \right ]'
  -
 \left [\psi_L^{(\ell)}(-a_{\ell})
  \right ]'=-i\xi_{\ell}\,\psi_L^{(\ell)}(-a_{\ell})
  , \ \ \
 \ell ={\cal L},{\cal L}-1,\ldots, 2,\\
 \left [
  \psi_C^{(0)}(-a_{1})
  \right ]'-
 \left [\psi_L^{(1)}(-a_{1})
  \right ]'=-i\xi_{1}\, \psi_C^{(0)}(-a_{1}),\ \ \\ \
 \left [
   \psi_R^{(1)}(a_1)
  \right ]'-
 \left [
  \psi_C^{(0)}(a_{1})
  \right ]'=i\xi_{1}\, \psi_C^{(0)}(a_{1}),\\
 \left [
  \psi_R^{(\ell+1)}(a_{\ell+1})
  \right ]'-
 \left [
  \psi_R^{(\ell)}(a_{\ell+1})
  \right ]'=i\xi_{\ell+1}\, \psi_R^{(\ell)}(a_{\ell+1}) , \ \ \
 \ell =1, 2, \ldots,{\cal L}-1,\\
 \ea
 \label{hladce}
 \ee
After the insertion of the ansatz (\ref{ansatz}), the set of
formulae (\ref{sesit}) and (\ref{hladce}) may be read as a
homogeneous linear algebraic system of $4{\cal L}$ equations for
the  $4{\cal L}$ unknown wave-function coefficients $\alpha_{\cal
L},\, \beta_{\cal L},\,\ldots,\,\nu$. The secular determinant
${\cal D}(\kappa)$ of this system must vanish so that the not too
complicated transcendental equation
 \be
 {\cal D}(\kappa)=0
 \ee
determines finally the set of the bound-state roots
$\kappa=\kappa_n$ at $n = 1, 2, \ldots$.

\subsection{Energies at the simplest choice of ${\cal L}=1$}

At ${\cal L}=1$, potential~(\ref{potik}) degenerates to the most
elementary double-well model with the single coupling $\xi_1=\xi$
and one displacement $a_1=a$ \cite{Kuba}. Out of the related eight
real constraints (\ref{sesit}) and (\ref{hladce}) only four are
independent and define the four real coefficients
$\alpha_1=\alpha$, $\beta_1=\beta$ and $\mu$ and $\nu$ as an
eigenvector of a four-by-four matrix with the secular determinant
 \be
 {\cal D}(\kappa)
 =
 -\frac{1}{2}
 \left \{
 \sin\, 2\kappa +
 \frac{\xi^2}{\kappa^2}\,
 \sin \,2\kappa a\,\cdot
 \sin^2 [\kappa(1- a)]
 \right \}
 \,.
 \label{sided}
 \ee
Numerically, the first term would give us the well-known
square-well spectrum at $\xi=0$, the completeness of which is
controlled by the Sturm-Liouville oscillation theory \cite{Ince}.
As long as all the roots $\kappa_n=\kappa_n(\xi)$ are smooth and
real functions of $\xi$ at the smallest couplings, $\kappa_n(\xi)
\approx n\pi/2 +{\cal O}(\xi^2/n)$, our explicit construction
confirms the general mathematical prediction \cite{Langer} that
the influence of the non-Hermiticity will be most pronounced at
the lowest part of the spectrum.

\subsection{The next choice of ${\cal L}=2$}

In the quadruple-well potential~(\ref{potik}) with ${\cal L}=2$ we
may shorten $a_1=a,\, a_2=b$ and drop the two redundant subscripts
in $\gamma_1=\gamma,\,\delta_1=\delta$. In the eight-dimensional
matrix of the system the elimination of four unknowns is either
trivial [$\gamma = \alpha_2\,\sin \kappa(1-b)$, $\delta =
\beta_2\,\sin \kappa(1-b)$] or easy [$\alpha_1=
\alpha_1(\alpha_2,\beta_2)$, $\beta_1=
\beta_1(\alpha_2,\beta_2)$]. We end up with a four-by-four matrix
problem and with the secular determinant
 \be
 {\cal D}(\kappa)
 =
 {\cal D}_{(0)}(\kappa)
 +
 {\cal D}_{(\xi_1)}(\kappa)
 +
 {\cal D}_{(\xi_2)}(\kappa)
 +
 {\cal D}_{(\xi_1\xi_2)}(\kappa),
  \label{sidekv}
 \ee
 \ben
 {\cal D}_{(0)}(\kappa)
 =
 -\frac{1}{2}
  \sin\, 2\kappa\, , \
  \ \ \ \ \ \ \ \ {\cal D}_{(\xi_j)}(\kappa)
 =
 -
 \frac{\xi_j^2}{2\kappa^2}\,
 \sin \,2\kappa a_j\,\cdot
 \sin^2 [\kappa(1- a_j)], \ \ \ \ j = 1, 2,
 \een
 \ben
 {\cal D}_{(\xi_1\xi_2)}(\kappa)
=
 -
 \left \{
 \frac{\xi_1\xi_2}{\kappa^2}\,
 \sin \,2\kappa a
 +
 \frac{\xi_1^2\xi_2^2}{\kappa^4}\,
 \sin^2 [\kappa(b-a)]\right \}
 \,
 \sin^2 [\kappa(1- b)]
 \,.
 \een
This secular determinant correctly degenerates to the previous
${\cal L} = 1$ formula in both the independent limits of $\xi_1
\to 0$ and $\xi_2 \to 0$.

\subsection{Simplifications at the rational $a_j$}

Let us return to the secular eq.~(\ref{sided}) with ${\cal L}=1$
and choose $a=1/2$ \cite{Kuba}. This leads to a factorization of
${\cal D}(\kappa)$ and to the pair of the eigenvalue conditions
 \be
  \cos {\kappa_{2m-1}} = \frac{\xi^2}{\xi^2-4\kappa_{2m-1}^2}, \ \
  \ \ \ \ \ \ \sin \kappa_{2m} = 0, \ \ \ \ \ m = 1, 2, \ldots
 \,
 \label{sinderela}
 \ee
with the second series of equations being exactly solvable, $
\kappa_{2m} = m\pi$.

At the next choice of $a=1/3$ we factorize  eq. (\ref{sided}) in
the similar manner and get the series of the $\xi-$dependent roots
specified by the implicit definitions
 \be
  \cos {\frac{4}{3}\kappa_{p}} =
  \frac{\xi^2+2\kappa_{p}^2}{\xi^2-4\kappa_{p}^2}, \ \ \
   \ \ \ \ \ p = 1, 2, 4,5,7,8,10,\ldots
 \,
 \label{derela}
 \ee
complemented by the closed formula for all the skipped roots of
the factor $\sin 2\kappa/3$ which remain $\xi-$independent and
read $ \kappa_{3m} =3 m\pi/2$ with $m = 1, 2, \ldots$. The
regularity of such a pattern of the $\xi-$independent roots is
easily prolonged to the decreasing sequence of $a$ with  $
\kappa_{4m} =2 m\pi$ at $a=1/4$ and all $m = 1, 2, \ldots$, etc.

The less elementary composite choice of $a=2/3$ may be observed to
give the same factor as at $a=1/3$ and, hence, the same
$\xi-$independent series of the roots $ \kappa_{3m} =3 m\pi/2$
with $m = 1, 2, \ldots$. The implicit formula for the remaining
roots is a slightly more complicated quadratic equation in the
trigonometric unknown $X=\cos 2\kappa/3$,
 \be
 \left (4\,\kappa^2 - \xi^2\right )\,X^2 + \xi^2\,X - \kappa^2=0
 \,.
 \ee
Its trigonometric part $X$ may be eliminated in the form
resembling eq.~(\ref{derela}).

One of the important consequences of the existence of the
elementary formulae for the rational $a$ is that they allow us to
perform an elementary analysis of the qualitative features of the
$n-$th root $\kappa_n$ during the growth of the strength $\xi$ of
the non-Hermiticity. During such an analysis one discovers that
these levels are either ``robust" (marked by a superscript,
$\kappa_n^{(R)}$, and remaining real for all $\xi$) or ``fragile"
(such a $\kappa_n^{(F)}$ will merge with another $\kappa_m^{(F)}$
at a ``critical" $\xi^{(C)}_{n,m}$ while the pair will complexify
beyind this ``exceptional" \cite{Heiss} point). For illustration
let us display this pattern in the three simplest spectra,
 \ben
 \kappa_1^{(F)},\
 \kappa_2^{(R)},\
 \kappa_3^{(F)},\
 \kappa_4^{(R)},\
 \kappa_5^{(F)},\
 \kappa_6^{(R)},\ \ldots, \ \ \ \ a = 1/2
 \een
 \ben
 \kappa_1^{(F)},\
 \kappa_2^{(F)},\
 \kappa_3^{(R)},\
 \kappa_4^{(F)},\
 \kappa_5^{(F)},\
 \kappa_6^{(R)},\ \ldots, \ \ \ \ a = 1/3
 \een
 \ben
 \kappa_1^{(F)},\
 \kappa_2^{(F)},\
 \kappa_3^{(R)},\
 \kappa_4^{(R)},\
 \kappa_5^{(R)},\
 \kappa_6^{(F)},\
 \kappa_7^{(F)},\
 \kappa_8^{(R)},\
 \kappa_9^{(R)},\
 \kappa_{10}^{(R)},\
 \kappa_{11}^{(F)},\ \ldots, \ \ \ \ a = 1/4.
 \een

%\newpage

\bigskip

 \noindent
{\small {\bf Acknowledgements}. Work partially supported by AS CR
(GA grant Nr. A1048302 and IRP AV0Z10480505). }

\bigskip

%\newpage

\end{document}